\documentclass[showpacs,amsmath,amssymb,prb,superscriptaddress,twocolumn]{revtex4}
\usepackage[T1]{fontenc}
\usepackage{amssymb,amsmath}
\usepackage{times}
\usepackage{graphicx}
\usepackage{wasysym}
\usepackage{subfigure}
\usepackage[sort&compress]{natbib}
\usepackage{bm}
\usepackage{overpic}
\usepackage{color}
\usepackage{multirow}

\begin{document}
\title{Conjugated gammadion chiral metamaterial with uniaxial optical activity and negative refractive index}

\author{R.~Zhao}
\affiliation{Ames Laboratory and Department of Physics and Astronomy,
             Iowa State University, Ames, Iowa 50011, USA}
\affiliation{Applied Optics Beijing Area Major Laboratory, Department of Physics,
Beijing Normal University, Beijing 100875, China}

\author{L.~Zhang}
\affiliation{Ames Laboratory and Department of Physics and Astronomy,
             Iowa State University, Ames, Iowa 50011, USA}
\author{J.~Zhou}
\affiliation{Centre for Integrated Nanotechnologies, Materials Physics \& Applications Division, Los Alamos National Laboratory, Los Alamos, New Mexico 87545, USA}

\author{Th.~Koschny}
\affiliation{Ames Laboratory and Department of Physics and Astronomy,
             Iowa State University, Ames, Iowa 50011, USA}
\affiliation{Institute of Electronic Structure and Laser, FORTH,
             Department of Materials Science and Technology, University of Crete, Heraklion, 71110 Crete, Greece}

\author{C.~M.~Soukoulis}
\affiliation{Ames Laboratory and Department of Physics and Astronomy,
             Iowa State University, Ames, Iowa 50011, USA}
\affiliation{Institute of Electronic Structure and Laser, FORTH,
             Department of Materials Science and Technology, University of Crete, Heraklion, 71110 Crete, Greece}
\date{\today}

\begin{abstract}
\noindent
We demonstrate numerically and experimentally a conjugated gammadion chiral metamaterial that uniaxially exhibits huge optical activity and circular dichroism, and gives a negative refractive index. This chiral design provides smaller unit cell size and larger chirality compared with other published planar designs. Experiments are performed at GHz frequencies (around 6\,GHz) and in good agreement with the numerical simulations.\\
\end{abstract}


\maketitle

\section{introduction}\label{introduction}
Materials whose magnetic/electric moment can be excited by the parallel external electric/magnetic field of the incident electromagnetic wave exhibiting optical activity \cite{Lindell1994} are called chiral materials. They are characterized by the quantity of \textit{chirality}, $\kappa=(n_{R}-n_{L})/2$, where $n_{R}$/$n_{L}$ is the refractive index of the right/left handed circular polarized wave (RCP/LCP). Natural chiral materials have very weak chirality (e.g., for quartz, $\kappa\simeq5\times10^{-5}$ at $\lambda=400 \,\mathrm{nm}$). Five orders of magnitude stronger chirality can be realized by chiral metamaterials made with sub-wavelength resonators. Chiral metamaterials  recently attracted a lot of interest because of strong chirality,\cite{A1} negative refractive index,\cite{Pendry2004, Tretyakov} and the prospect of a repulsive Casimir force.\cite{Zhao2009} Many chiral metamaterial designs have been proposed and demonstrated to obtain large optical activity, circular dichroism,\cite{Decker2009,Decker2010} and negative refractive index.\cite{Plum2009, Zhang2009, Zhou2009, Wang2009, Wang2009JOA}

In this article, we study both numerically and experimentally the conjugated gammadion chiral metamaterial that uniaxially exhibits huge optical activity, circular dichroism, and negative refractive index. This new design more clearly lacks any mirror symmetry plane compared to previous work.\cite{Decker2007} In Ref. [13], the structure itself has a mirror symmetry plane. The lack of mirror symmetry only originates from the presence of the substrate and different trace widths in the bi-layer structure due to fabrication constraints. While in our new design, the trace widths are the same and the lack of mirror symmetry is an inherent property of the particular design. And meanwhile, our new design possesses smaller unit cell size and larger chirality compared with other published planar designs: twisted-rosettes,\cite{Plum2009} twisted-crosswires,\cite{Zhou2009,Decker2009} and four-U-SRRs.\cite{Decker2010, ZhaofengAPL2010, XiongPRB2010}.

\section{Experimental and simulation}\label{Experimental and simulation}
\begin{figure}[htb!]
\centerline{\includegraphics[width=0.48\textwidth]{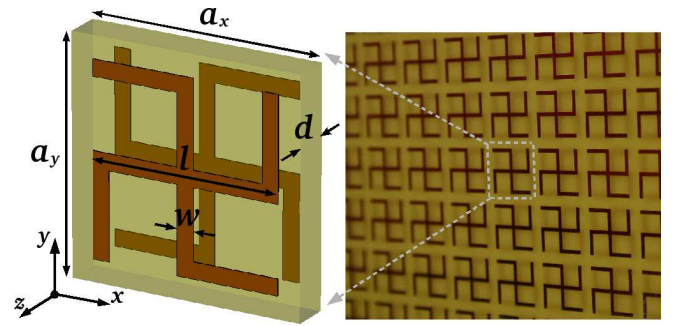}}
\caption{(Color online) Scheme of the conjugated gammadion chiral metamaterial. Two gammadion copper resonators are conjugatedly aligned on each side of the FR-4 board. The geometry parameters are given as $a_x=a_y=10\,\mathrm{mm}$, $l=8.1\,\mathrm{mm}$, $w=0.7\,\mathrm{mm}$, $d=1.6\,\mathrm{mm}$, and the thickness of copper is $t=0.036\,\mathrm{mm}$.\label{structure}}
\end{figure}

The layout of the conjugated gammadion chiral metamaterial is shown in Fig. 1. A $30\times30$ array of the conjugated gammadion resonator pairs is patterned on each side of the FR-4 board. The relative dielectric constant of the FR-4 board is $\epsilon=4.2$ with a dielectric loss tangent of 0.02. The two layers of the gammadion resonators are conjugatedly arranged in order to break the mirror symmetry along the direction perpendicular to the metamaterial plane. The metamaterial possesses $C_4$ symmetry on the $z$-axis and, therefore, exhibits uniaxial chirality for the normal incident electromagnetic wave. The dimensions of the unit cell and the photograph of the experimental sample are shown in Fig. {\ref{structure}}.

The amplitudes and phases of the linearly polarized transmissions, $T_\parallel$ (polarization of the transmitted wave is parallel with that of the incident wave) and $T_\perp$ (polarization of the transmitted wave is perpendicular to that of the incident wave), and reflection coefficient $R$ ($R=R_\parallel, R_\perp=0$)\cite{Lakhtakia} are measured using an HP 8364B network analyzer with two Narda standard horn antennas. The circular polarized transmissions of the RCP and LCP, $T_R$ and $T_L$, and reflections, $R_L$ and $R_R$, can be obtained from the linearly polarized transmission coefficients by:
\begin{equation}
T_{R}=T_\parallel-iT_\perp, T_{L}=T_\parallel+iT_\perp, R_{R}=R_{L}=R. \label{coefficient}
\end{equation}
The circular polarized waves are the eigenwave functions of the chiral metamaterials. The cross transmissions, from LCP to RCP and vice versa, are zero. The numerical simulations were performed using the frequency domain solver of the CST Microwave Studio (Computer Simulation Technology GmbH, Darmstadt, Germany), which implements a finite element method. In the simulations, the unit cell boundary condition was applied and the circular polarized eigenwaves used directly.

\section{Results}\label{Results}
Figures. {\ref{transmissionandreflection}}(a) and (b) show the simulation (left) and experimental (right) results of the transmissions and reflections. The calculated results nicely agree with our experimental results. There are two resonances in the transmission spectrum. One corresponds to a peak around $f=5.6\,\mathrm{GHz}$ and the other, a dip, around $f=7.8\,\mathrm{GHz}$.  The first resonance is much sharper than the second one. The feature of the transmission peak for both RCP and LCP at the first resonance is observed for the first time in chiral metamaterials. From the following retrieval results, we conclude that this peak is induced by the strong magnetic response, which decreases the degree of impedance mismatch. At the resonances, the transmission spectra for the RCP and LCP waves are significantly different. At the first resonance, the transmission of LCP is larger than that of the RCP; at the second resonance, it reverses. This difference between the amplitudes of two transmissions is characterized by ellipticity, $\eta=\frac{1}{2}\tan^{-1}(\frac{|T_L|^2-|T_R|^2}{|T_L|^2+|T_R|^2})$, shown in Figs. {\ref{transmissionandreflection}(e) and (f). Far from the resonances, the phases of the RCP and LCP converge. In the vicinity of the resonances, they are obviously different, which will induce the rotation of the polarization plane of a linearly polarized light as it passes through the chiral metamaterial. The difference between the phases is characterized by the rotary power, $\theta=\frac{1}{2}[\arg(T_L)-\arg(T_R)]$, shown in Figs. {\ref{transmissionandreflection}(c) and(d). The most interesting region is around $\eta=0$, which will give the "pure" optical activity effect, i.e., for the linearly polarized incident wave, the transmission wave will still be linearly polarized just with the polarization plane rotated by $\theta$ degree. This design possesses very strong optical activity. At $\eta=0$ (i.e., $f=6.55\,\mathrm{GHz}$), the rotation angle is about 30 degrees for 1.6\,mm thick metamaterial, i.e., the rotation angle is as large as $856^\circ/\lambda$.
%
\begin{figure}[htb!]
\centerline{\includegraphics[width=0.48\textwidth]{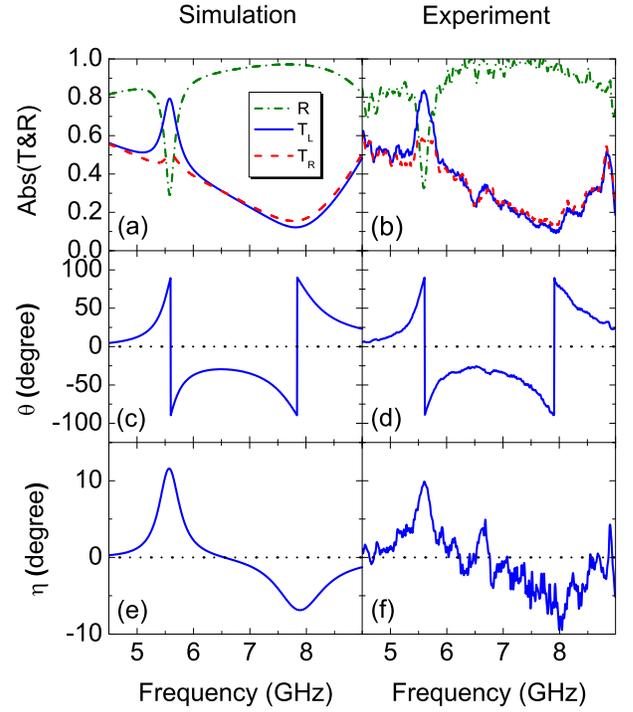}}
\caption{(Color online) The simulation (left) and experimental (right) results of the transmissions and reflections, and the rotation angle and circular dichroism. \label{transmissionandreflection}}
\end{figure}
\begin{figure}[htb!]
\centerline{\includegraphics[width=0.48\textwidth]{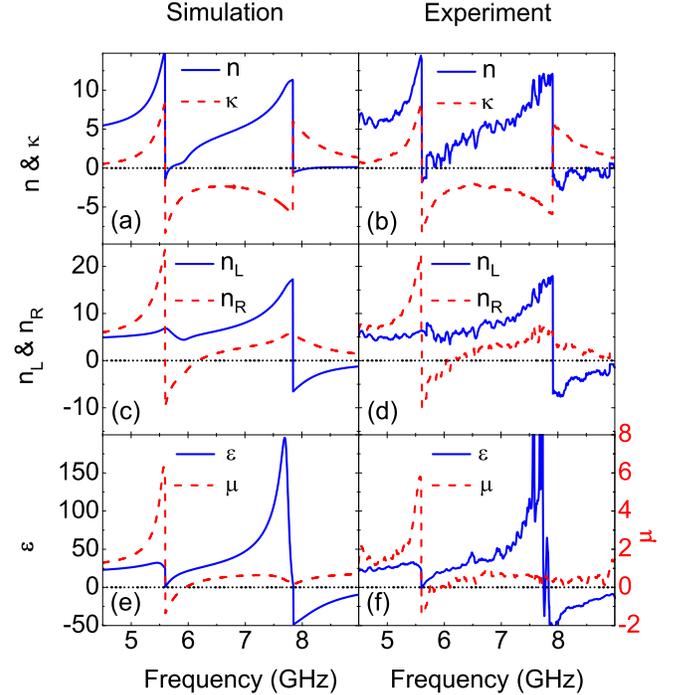}}
\caption{(Color online) The retrieval results from the numerical (left) and experimental (right) results of the transmissions and reflections. \label{retrieval}}
\end{figure}

After obtaining the data of the transmissions and reflections, we can apply a retrieval procedure to obtain the effective constitutive parameters. \cite{Zhao2010oe, Smith2002} The chirality $\kappa$ can be obtained directly from the transmissions as
\begin{subequations}
\label{kapparealimag}
\begin{align}
\mathrm{Re}(\kappa)&= \frac{\arg(T_L)-\arg(T_R)+2m\pi}{2k_0d}, \\
\mathrm{Im}(\kappa)&= \frac{\ln|T_L|-\ln|T_R|}{2k_0d},
\end{align}
\end{subequations}
where $k_0$ is the wave vector in the vacuum; $d$ is the thickness of the sample (here, the thickness of the retrieved sample is chosen as 1.6\, mm, i.e., the thickness of the FR-4 board) and the integer, $m$, is determined by the condition of $-\pi<\arg(T_L)-\arg(T_R)+2m\pi<\pi$ for one unit cell. In previous studies,\cite{Jiangfeng2008,Jiangfengcoupling} the retrieved parameters depend on the thickness of the sample \cite{Jiangfeng2008} and once we have multi-layer sample, the retrieved parameters, for the weakly coupled system, are very close to the one unit cell and converge very fast \cite{Jiangfengcoupling}. For the strongly coupled system, the retrieved parameters are completely different from the one unit cell system.\cite{Jiangfengcoupling, Zhaofeng}

Comparing with Figs. \ref{transmissionandreflection}(c-f), we note that the real and imaginary parts of $\kappa$ relate to the azimuth rotation angle, $\theta$, and the circular dichromism, $\eta$, respectively. The average refractive index, $n=(n_L+n_R)/2$, and the impedance $z$ can be obtained by using the traditional retrieval procedure \cite{Smith2002} after taking the geometric average transmission, $T=\sqrt{T_LT_R}$. The branch of the square root is chosen coordinating with the real part of $\kappa$. Then, the other parameters can be calculated by: $n_\pm=n\pm\kappa$, $\epsilon=n/z$, and $\mu=nz$. Figure \ref{retrieval} shows the retrieval results. The chirality is very large, $\kappa=2.35$ at $\eta=0$, which corresponds to frequency, $f=6.55\,\textrm{GHz}$ ($\kappa=2.09$ for the experimental result). This strong chirality can easily push the refractive index of the RCP/LCP eigenwave, $n_\pm=n\pm\kappa$, to become negative at the resonances as shown in Fig. \ref{retrieval}(c) and (d). Due to the loss of the PC board, the figure of merit (FOM), $-\mathrm{Re}(n)/\mathrm{Im}(n)$, is very low compared with the traditional negative index metamaterials.\cite{Soukoulis2008} If using a lower loss substrate, the FOM can be larger than $10$ according to our numerical simulation. In Figs. \ref{retrieval}(e) and (f), we show the retrieved real parts of the permittivity and permeability of the chiral metamaterials.

\section{discussion}\label{discussion}

To illustrate the origin of the chiral response of our metamaterial, in Fig. \ref{currentdistrubution}, we discuss a procedure of transmutation from the simple $\Omega$-particle chiral element to the conjugated gammadion chiral metamaterial. The $\Omega$-particle chiral element is one of the important chiral structures studied analytically elsewhere.\cite{Lindell1994} Here, we place two $\Omega$-particle chiral elements together to form a conjugated $\Omega$-particle chiral element pair. The linearly polarized external electric field can drive the electric dipole via the vertical arms. This electric response (ER) generates circular electric current on the loop which gives a magnetic moment in the same direction as the electric dipole. Rotating the conjugated $\Omega$-particle chiral element pair by 90 degrees, the external magnetic field will drive the magnetic moment via the loop. This magnetic response (MR) then drives the current on the vertical arms which gives an electric dipole in the same direction as the magnetic moment. This clearly shows how the magnetic/electric moment is induced by the electric/magnetic field (of the incident EM wave) in the parallel direction. According to the aforementioned definition, this results in the chirality. If adding the two $\Omega$-particle pairs together, it will be a bi-isotropic uniaxial chiral metamaterial. After stretching the semi-circle loop to a straight line, it becomes to our current planar design.
\begin{figure}[htb!]
\centerline{\includegraphics[width=0.43\textwidth]{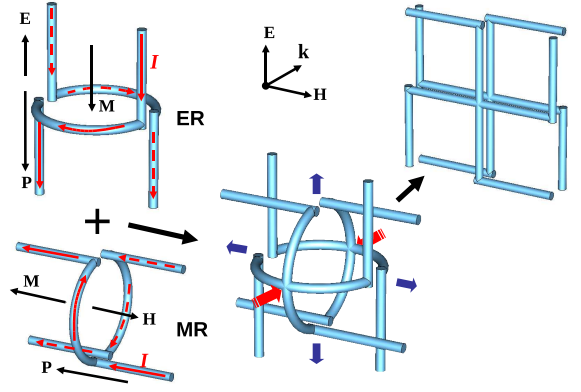}}
\caption{(Color online) The procedure of the transmutation from the simple $\Omega$-particle chiral element to the conjugated gammadion chiral metamaterial. The procedure shows how the magnetic/electric moment is induced by the electric/magnetic field (of the incident EM wave) in the parallel direction. ER and MR are the electric and magnetic response modes, respectively. The arrow indicates the current direction. \label{currentdistrubution}}
\end{figure}

\begin{figure}[htb!]
\begin{overpic}[width=0.45\textwidth]{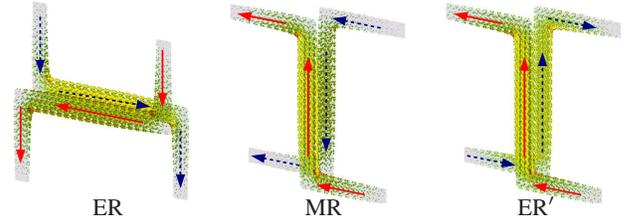}
\put(13,-2){ER}
\put(48,-2){MR}
\put(83,-2){ER$^\prime$}
\end{overpic}
\caption{(Color online) Three basic electromagnetic responses in conjugated gammadion chiral metamaterials. The arrow indicates the current direction. ER and MR are comparable with those in Fig. \ref{currentdistrubution}. ER$^\prime$ is a new emerged electric response which is driven by the central arms. ER$^\prime$ can not give chirality response. The incident wave is the same as in Fig. \ref{currentdistrubution}. \label{currentdistrubution2}}
\end{figure}
\begin{figure}[htb!]
\begin{overpic}[width=0.3\textwidth]{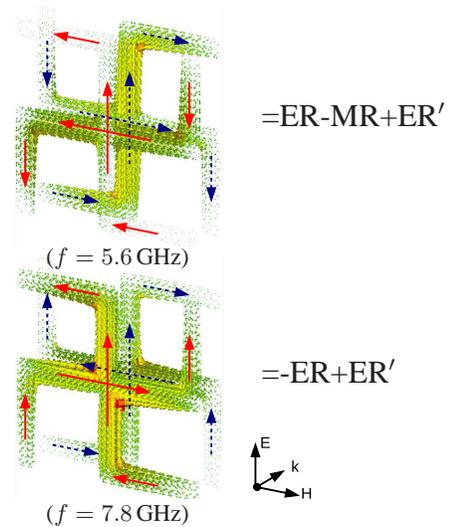}
\put(8,48){($f=5.6$\,GHz)}
\put(8,-2){($f=7.8$\,GHz)}
\put(50,75){\large =ER-MR+ER$^\prime$}
\put(50,25){\large =-ER+ER$^\prime$}
\end{overpic}
\caption{(Color online) Illustration of superposition at $f=5.6$\,GHz (upper) and $f=7.8$\,GHz (down). \label{currentmode}}
\end{figure}

To further illustrate the origin of chirality in our conjugated gammadion chiral metamaterials, Fig. \ref{currentdistrubution2} shows three basic electromagnetic responses (ER, MR, and ER$^\prime$) in conjugated gammadion chiral metamaterials. ER and MR are comparable with those in Fig. \ref{currentdistrubution}. ER$^\prime$ is an additional, non-resonant electric response which is driven by the central parallel arms. ER$^\prime$ can not give chirality response because there is no magnetic moment generated. In our conjugated gammadion chiral metamaterials, each resonance is a superposition of these three electromagnetic response modes. For example, as shown in Fig. \ref{currentmode}, the resonance at f=$5.6$\,GHz can be considered as the superposition of ER, MR, and ER$^\prime$. The existence of the MR is also reflected by the retrieval results in Figs. \ref{retrieval}(e) and \ref{retrieval}(f), where the effective permeability has a strong resonance at f=$5.6$\,GHz. And the resonance at f=$7.8$\,GHz can be considered as the superposition of ER and ER$^\prime$, then the retrieval results only show the electric response.

\begin{figure}[htb!]
\begin{overpic}[width=0.48\textwidth]{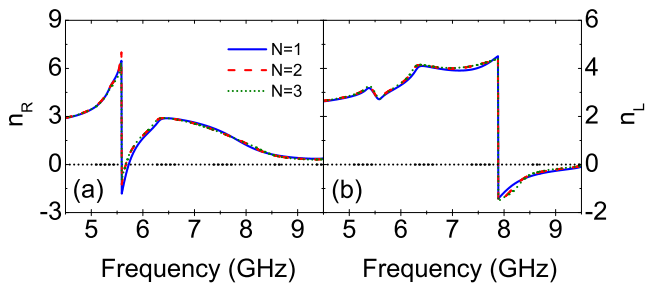}
\end{overpic}
\caption{(Color online) The retrieved refractive indices for multi-layer conjugated gammadion chiral metamaterial \label{multi-unit}}
\end{figure}
Some may raise the question whether our design is just a thin meta-surface that exhibits rotation of polarization phase or can represent the properties of the bulk chiral metamaterial as we expected. In order to clarify this point, in Fig. \ref{multi-unit} we plot the retrieved refractive indices for multi-layer metamaterial (a pair of doule-layered conjugated gammadions is called one layer.) . Here, we study the weakly coupling case:\cite{Jiangfengcoupling} The separation $D$ between each layer of conjugated gammadion pairs is three times of the thickness of FR-4 board, i.e., $D=3d$ and the total thickness of each unit cell is $D+d=6.4\textrm{mm}$. Figure \ref{multi-unit} shows that the retrieved refractive indices for the multi-layer metamaterial is rapidly converged. Three layers (N=3) almost overlaps with two layers (N=2). Therefore, the single layer we studied previously can represent the properties of bulk chiral metamaterials for these weakly coupling cases. The differences between the refractive indices in Fig. \ref{retrieval} and Fig. \ref{multi-unit} for the single unit cell are due to the different thicknesses of the unit cell. In Fig. \ref{retrieval}, we take the unit cell thickness as 1.6 mm, while it's 6.4 mm in Fig. \ref{multi-unit}. Comparing Fig. \ref{multi-unit} and Figs. \ref{retrieval}(e) and \ref{retrieval}(f), the magnitude of the retrieved effective parameters of $n_R$ and $n_L$ decrease, as the size of the unit cell increases. \cite{Jiangfeng2008}   Note, for the strong coupling cases (i.e., small $D$), the strong coupling between each layer will induce differences between one layer and multi-layer structure \cite{Zhaofeng} and slow the convergence of the retrieved parameters.\cite{Jiangfengcoupling}

\

\section{conclusion}\label{conclusion}
In conclusion, we have designed and studied a conjugated gammadion chiral metamaterial at around 6\,GHz. This chiral metamaterial exhibits huge uniaxial optical activity. The rotation angle is as large as $30^\circ$ for 1.6\,mm thick metamaterial, corresponding to $\kappa=2.35$ with $\eta=0$. The chiral metamaterial also exhibits uniaxial negative refractive index, due to its strong chirality. The origin of the chirality is intuitionally and physically explained, based on the $\Omega$-particle chiral element model.

\section{ACKNOWLEDGMENT}\label{ACKNOWLEDGMENT}
Work at Ames Laboratory was supported by the Department of Energy
(Basic Energy Sciences) under contract No.~DE-AC02-07CH11358. This
work was partially supported by the European Community FET project
PHOME (Contract No.~213390) and by the Department of Navy, Office of Naval Research (Grant No. N000141010925). R. Z. acknowledges the China Scholarship Council (CSC) for financial support.


\end{document}